\documentclass[epj]{svjour}
\usepackage{graphicx}
\usepackage{amsmath,amsfonts}

\newcommand{\re}{{\rm e}}
\newcommand{\ri}{{\rm i}}

\begin{document}
\title{Uniform semiclassical approximations on a topologically non-\\trivial
       configuration space: The hydrogen atom in an electric field}
\titlerunning{Uniform semiclassical approximations on a topologically non-trivial
       configuration space}
\author{Thomas Bartsch \and J\"org Main \and G\"unter Wunner}

\institute{Institut f\"ur Theoretische Physik 1, Universit\"at Stuttgart,
           D-70550 Stuttgart, Germany}

\date{Received: date / Revised version: date}

\abstract{
Semiclassical periodic-orbit theory and closed-orbit theory
represent a quantum spectrum as a superposition of contributions from
individual classical orbits.  Close to a bifurcation, these contributions
diverge and have to be replaced with a uniform approximation.  Its
construction requires a normal form that provides a local description of
the bifurcation scenario.  Usually, the normal form is constructed in flat
space. We present an example taken from the hydrogen atom in an electric
field where the normal form must be chosen to be defined on a sphere
instead of a Euclidean plane. In the example, the necessity to base the
normal form on a topologically non-trivial configuration space reveals a
subtle interplay between local and global aspects of the phase space
structure.  We show that a uniform approximation for a bifurcation scenario
with non-trivial topology can be constructed using the established
uniformization techniques.  Semiclassical photo-absorption spectra of the
hydrogen atom in an electric field are significantly improved when based on
the extended uniform approximations.
}


\maketitle

\section{Introduction}
\label{sec:intro}

Periodic-orbit theory \cite{Gutzwiller90,Brack97} and, as a variant for
atomic photo-absorption spectra, closed-orbit theory
\cite{Du88,Bogomolny89} have proved to be the most powerful tools available
for a semiclassical interpretation of quantum mechanical spectra.
Closed-orbit theory provides a semiclassical approximation to the quantum
response function
\begin{equation}
  \label{gDef}
  g(E)=-\frac 1{\pi}
       \sum_n \frac{|\left<i|D|n\right>|^2}{E-E_n+\ri\epsilon} \;,
\end{equation}
which involves the energy eigenvalues $E_n$ and the dipole matrix elements
$\left<i|D|n\right>$ connecting the excited states $n$ to the initial state
$i$.
Semiclassically, the response function splits into a smooth part and an
oscillatory part of the form
\begin{equation}
  \label{resOscGen}
  g^{\rm osc}(E) =
    \sum_{\rm c.o.} {\cal A}_{\rm c.o.}(E)\,\re^{\ri S_{\rm c.o.}(E)/\hbar} \;,
\end{equation}
where the sum extends over all classical closed orbits starting from the
nucleus and returning to it after having been deflected by the external fields.
$S_{\rm c.o.}$ is the classical action of the closed orbit. The precise
form of the recurrence
amplitude $\cal A_{\rm c.o.}$ depends on the geometry of the external
field configuration, but it can in any case be calculated from classical
properties of the closed orbit. Periodic-orbit theory, which can be applied
to any quantum system possessing a classical counterpart, represents the
quantum density of states as a sum over the classical periodic orbits in
complete analogy with~(\ref{resOscGen}).

Both variants of the semiclassical theory share the problem that the
semiclassical amplitudes ${\cal A}_{\rm c.o.}$ diverge when the
corresponding classical orbits undergo a bifurcation. This failure occurs
because close to a bifurcation the classical orbits cannot be regarded as
isolated in phase space, as the simple semiclassical
expansion~(\ref{resOscGen}) assumes. It can be remedied if the individual
contributions of the bifurcating orbits to the closed-orbit
sum~(\ref{resOscGen}) are replaced with a uniform approximation that
describes their collective contribution
\cite{Ozorio87,Sieber96,Schomerus97b,Sieber98}. For the generic
codimension-one bifurcations of periodic orbits
\cite{Sieber96,Schomerus97b,Sieber98} as well as closed orbits
\cite{Bartsch02,Bartsch03b}, uniform approximations are available in the
literature. Typically, bifurcations of higher codimensions are not
encountered classically because they split into sequences of
codimension-one bifurcations upon an arbitrarily small
perturbation. Semiclassically, however, these sequences must be described
collectively by a single uniform approximation if the individual
bifurcations are sufficiently close
\cite{Main97a,Schomerus97a,Main98a,Schomerus98}. It can thus be necessary
in practice to construct uniform approximations for bifurcation scenarios
that are much more complicated than the elementary bifurcations of
codimension one.

The crucial step in the derivation of a uniform approximation
\cite{Sieber98,Bartsch02,Bartsch03b,Main97a} is the choice of a normal form
$\Phi_a(x)$ describing the bifurcation scenario. Its stationary points,
i.e. the solutions of $\partial\Phi_a(x)/\partial x=0$, must be in
one-to-one correspondence with the classical closed orbits. In addition, as
the parameter $a$ is varied, the stationary points must collide and change
from real to complex in accordance with the closed-orbit bifurcations. Once
a suitable normal form is known, the uniform approximation assumes the form
\begin{equation}
  \label{StarkUnif}
  \Psi(E) = I(a)\,\re^{\ri S_{0}(E)/\hbar} \;,
\end{equation}
where
\begin{equation}
  \label{StarkInt}
  I(a) = \int dx\, p(x)\,\re^{\ri\Phi_a(x)}
\end{equation}
and the normal form parameter $a$, the action $S_0$ and the unknown
function $p(x)$ are determined as functions of the energy $E$ such that the
uniform approximation asymptotically reproduces the closed-orbit
formula~(\ref{resOscGen}) for isolated orbits if the distance from the
bifurcation is large.

Usually, the choices of $a$ and $S_0$ are unique. By contrast, the
asymptotic condition fixes the values of the amplitude function $p(x)$ only at
the stationary points of $\Phi_a$, so that there remains considerable
freedom in its choice. Customarily, a low-order polynomial that
interpolates between the given values is chosen. This decision can be
justified by observing that a uniform approximation is needed only in the
vicinity of the bifurcation, where the stationary points of the normal form
are close. Since, in the spirit of the stationary-phase approximation, only
the neighbourhood of the stationary points of $\Phi_a$ contributes to the
integral~(\ref{StarkInt}), a Taylor series approximation that is accurate
in that region can be expected to be appropriate.

So far, the configuration space of a uniform approximation, i.e.~the domain
of the normal form variable $x$ and the range of the
integral~(\ref{StarkInt}) could always be chosen to be a one- or
multidimensional Euclidean space. In this paper, we describe an example
where this is impossible because for topological reasons the bifurcating
orbits cannot be represented as stationary points of a function defined on
a Euclidean space. This result is surprising because uniform
approximations, as well as the underlying normal forms, are needed to
describe closely spaced orbits and bifurcations, and in this sense are local
constructions. The topology of the configuration space, on the other hand,
reflects its global characteristics. The necessity to base a uniform
approximation on a configuration space of a suitable topology thus involves
a subtle interplay between local and global aspects of the classical phase
space structure.  We demonstrate that the established techniques for the
construction of uniform approximations are, in principle, applicable to
uniform approximations on a topologically non-trivial configuration space,
but that additional complications arise when non-local aspects of the
dynamics have to be dealt with.

The bifurcation scenario to be discussed arises, e.g., in the hydrogen atom
in an electric field.  This is an integrable system, and photo-absorption
spectra can in principle be obtained by the quantization of classical tori
\cite{Kon97,Kon98}.  However, while the torus quantization is restricted to
integrable systems, closed-orbit theory is intended to be applicable to
arbitrary atomic systems exhibiting either regular, chaotic, or mixed
classical behaviour. Therefore, it is of vital interest to see how it can
be applied to this apparently simple example. It turns out that to a
closed-orbit theory approach the hydrogen atom in an electric field poses
the same difficulties in principle as a generic mixed system because the
closed orbits undergo plenty of bifurcations. We thus start with a
description of the closed orbits in that system in
section~\ref{sec:StarkClosed}.  In section~\ref{sec:StarkLow} we focus on
the bifurcations at low energies and show that in this spectral region a
uniform approximation is required that goes beyond the known uniform
approximation for the Stark system
\cite{Gao97,Shaw98a,Shaw98b,Bartsch02a,Bartsch03d}. This uniform
approximation is constructed in section~\ref{sec:StarkUniform}, where we
also demonstrate that it must rely on a sphere instead of a Euclidean plane
as its configuration space. It is compared with the previous uniform
approximation, and in section~\ref{sec:StarkScl} its merits are assessed by
the calculation of semiclassical photo-absorption spectra.

\section{Closed orbits}
\label{sec:StarkClosed}

The bifurcations of closed orbits in the hydrogen atom in an electric field
follow a simple scheme that was described by Gao and Delos
\cite{Gao94}. Complementary to that work, we demonstrated in a recent
publication \cite{Bartsch03d} that to a large extent the closed orbits and
their parameters can be computed analytically. We will now summarize the
essential features of the bifurcation scenario. For details, the reader is
referred to \cite{Bartsch03d}.

If we assume the electric field $\vec F$ to be directed along the $z$-axis,
the Hamiltonian describing the motion of the atomic electron reads, in
atomic units,
\begin{equation}
  \label{classHam}
  H = \frac{1}{2}\vec p^2 -\frac 1r + F z \;,
\end{equation}
where $r^2=x^2+y^2+z^2$. It can be simplified by means of its scaling
properties: If the coordinates and momenta are scaled according to
$\tilde{\vec x}=F^{1/2}\vec x$, $\tilde{\vec p}=F^{-1/4}\vec p$, the classical
dynamics is found not to depend on the energy $E$ and the electric field
strength $F$ separately, but only on the scaled energy $\tilde
E=F^{-1/2}E$.

The qualitative simplicity of the bifurcation scenario is due to the
well-known fact \cite{Born25,LandauI} that the Hamiltonian~(\ref{classHam})
is separable in semiparabolic coordinates $\rho_1$, $\rho_2$, $\varphi$,
where
\begin{equation}
  \rho_1 = \sqrt{\tilde r-\tilde z}\;, \qquad
  \rho_2 = \sqrt{\tilde r+\tilde z}\;,
\end{equation}
and $\varphi$ is the azimuth angle around the electric field axis. The
angular coordinate $\varphi$ is ignorable, so that its conjugate momentum,
the angular momentum $L_z$ around the field axis, is conserved. For closed
orbits, which pass through the nucleus, $L_z=0$. The dynamics of $\rho_1$
and $\rho_2$ is then given by the equations of motion
\begin{equation}
  \begin{split}
    \rho_1'^2 = \phantom{+} \rho_1^4 + 2 E \rho_1^2 + c_1 \;,\\
    \rho_2'^2 = - \rho_2^4 + 2 E \rho_2^2 + c_2 \;,\\
  \end{split}
\end{equation}
where $E$ is the energy and the separation constants $c_1$ and $c_2$
describe the distribution of energy between the $\rho_1$ and $\rho_2$
degrees of freedom. They are restricted by
\begin{equation}
  \label{cConst}
  c_1 + c_2 = 4 \;.
\end{equation}
For real orbits the separation constants are constrained to the interval
\begin{equation}
  \label{cInt}
  0 \le c_j \le 4 \;.
\end{equation}
If they fail to meet~(\ref{cInt}), either $\rho_1$ or
$\rho_2$ assumes complex values, so that a closed ``ghost'' orbit in the
complexified phase space is obtained.

In the simplest cases, one of the $\rho_j$ vanishes identically. The
electron then moves along the electric field axis.  If $c_1=0$, the
downhill coordinate $\rho_1$ is zero. The electron leaves the nucleus in
the uphill direction, i.e. in the direction of the electric field, until it
is turned around by the joint action of the Coulomb and external fields and
returns to the nucleus. After the first return, the uphill orbit is
repeated periodically. As the external field provides an arbitrarily high
deflecting potential, the uphill orbit is closed at all energies.

The second case of axial motion is obtained if $c_1=4$, which corresponds
to $\rho_2\equiv 0$ and thus to a downhill motion opposite to the direction
of the electric field.  The downhill orbit closes only if the scaled energy
$\tilde E$ is less than the scaled saddle point energy $\tilde E_{\rm
S}=-2$. If $\tilde E>\tilde E_{\rm S}$, the electron escapes to infinity
and the orbit does not close. At scaled energies $\tilde E<\tilde E_{\rm
S}$, the downhill orbit, as the uphill orbit, is repeated periodically.

In general, the motion of the electron will not be directed along the
electric field axis and will consist of a superposition of the downhill
and uphill motions. In this case a closed orbit is obtained if the
downhill and uphill motions close at the same time after $k$ oscillations
in the downhill direction and $l$ oscillations in the uphill direction. A
closed orbit characterized by integer repetition number $(k,l)$ must
satisfy \cite{Bartsch03d}
\begin{equation}
  \label{ClosedCond}
  k\rho_-\,{\cal K}(m_1) = l\rho_+\,{\cal K}(m_2)
\end{equation}
with
\begin{equation}
\begin{alignedat}{2}
   \rho_+&=\sqrt{-\tilde E+\sqrt{\tilde E^2-c_1}} \;, \qquad & 
   m_1 &= +\frac{c_1}{\rho_+^4} \;,\\
   \rho_-&=\sqrt{-\tilde E+\sqrt{\tilde E^2+c_2}} \;, \qquad&
   m_2 &= -\frac{c_2}{\rho_-^4}
\end{alignedat}
\end{equation}
and ${\cal K}(m)$ denoting the complete elliptic integral of the first kind
\cite{Abramowitz}.

For given repetition numbers $k,l$ and scaled energy $\tilde E$, if we use
that $c_2=4-c_1$ by~(\ref{cConst}), equation~(\ref{ClosedCond}) can be
solved for the separation parameter $c_1$ that specifies the initial
conditions of the orbit. A solution only exists if $l>k$. Once $c_1$ is
known, all orbital parameters relevant to closed-orbit theory can be found
analytically. In particular, the classical action of the closed orbit is
\cite{Bartsch03d}
\begin{equation}
  \label{StarkAction2}
    S = \frac{2}{3}\,k\rho_+^3I(m_1) -
        \frac{2}{3}\,l\rho_-^3I(m_2)
\end{equation}
with
\begin{equation}
  I(m)=(m-1){\cal K}(m) + (m+1){\cal E}(m)
\end{equation}
and the complete elliptic integrals of the first and second kinds ${\cal
K}(m)$ and ${\cal E}(m)$ \cite{Abramowitz}. It is real for both real and
ghost orbits.

If the repetition numbers $k$ and $l$ are fixed, for sufficiently low
scaled energies $\tilde E$ the value of the separation constant $c_1$
obtained from~(\ref{ClosedCond}) is greater than 4, so that the
$(k,l)$-orbit is a ghost. As $\tilde E$ increases, $c_1$ decreases
monotonically. At the energy $\tilde E_{\rm dest}$ where $c_1=4$, the
$(k,l)$-orbit becomes real. It is generated (as a real orbit) in a
bifurcation off the $k$-th repetition of the downhill orbit, which is
characterized by $c_1=4$. As $\tilde E$ is further increased, the orbit
moves away from the downhill orbit and approaches the uphill orbit. At the
energy $\tilde E_{\rm dest}$ where $c_1=0$, the orbit $(k,l)$ collides with
the $l$-th repetition of the uphill orbit and is destroyed. At $\tilde
E>\tilde E_{\rm dest}$ the orbit is a ghost again.

The bifurcation energies $\tilde E_{\rm gen}$ and $\tilde E_{\rm dest}$ can
be determined from~(\ref{ClosedCond}) if $c_1=4$ is prescribed for the
generation of an orbit, or $c_1=0$ for its destruction. For these cases,
equation~(\ref{ClosedCond}) simplifies to
\begin{equation}
  \label{genEq}
  \frac{l}{k} = \frac{2^{3/2}}{\pi\sqrt{1+\sqrt{1-\epsilon}}}
    \,\,{\cal K}\left(\frac{\epsilon}
                           {\left(1+\sqrt{1-\epsilon}\right)^2}\right)
\end{equation}
in terms of the dimensionless variable $\epsilon=4/\tilde E^2=(\tilde
E_{\rm S}/\tilde E)^2$ for the generation of the $(k,l)$ orbit and
\begin{equation}
  \label{destEq}
  \frac{k}{l} = \frac{2^{3/2}}{\pi\sqrt{1+\sqrt{1+\epsilon}}}
    \,\,{\cal K}\left(\frac{-\epsilon}
                           {\left(1+\sqrt{1+\epsilon}\right)^2}\right)
\end{equation}
for its destruction.

\section{Bifurcations at low energies}
\label{sec:StarkLow}

From the discussion of the closed orbit bifurcations in the preceding
section it is evident that there is only a single elementary bifurcation
scenario in the hydrogen atom in an electric field, viz.~a non-axial orbit
being born or destroyed in a collision with an axial orbit. Uniform
semiclassical approximations describing this type of bifurcation have been
derived by several authors
\cite{Gao97,Shaw98a,Shaw98b,Bartsch03d}. However, they are only applicable
if the individual bifurcations are sufficiently isolated. If they are
close, more complicated uniform approximations that provide a collective
description of several bifurcations must be found. Thus, there are two
spectral regions where the construction of uniform approximations poses a
particular challenge:

First, as the Stark saddle energy $\tilde E_{\rm S}=-2$ is approached from
below, the downhill orbit undergoes infinitely many bifurcations in a
finite energy interval before it is destroyed. The periods of the closed
orbits  created in this way, as the period of the downhill orbit itself, grow
arbitrarily large. The actions, however, remain finite, so that the action
differences between the orbits stemming from two successive bifurcations
become arbitrarily small. Similar cascades of isochronous bifurcations have
been found, e.g., in the diamagnetic Kepler problem \cite{Wintgen87} and in
H\'enon-Heiles type potentials \cite{Brack01}.  Their uniformization
remains an open problem. A detailed discussion of this cascade in the Stark
system must be referred to future work.

\begin{figure}
  \centerline{\includegraphics[width=\columnwidth]{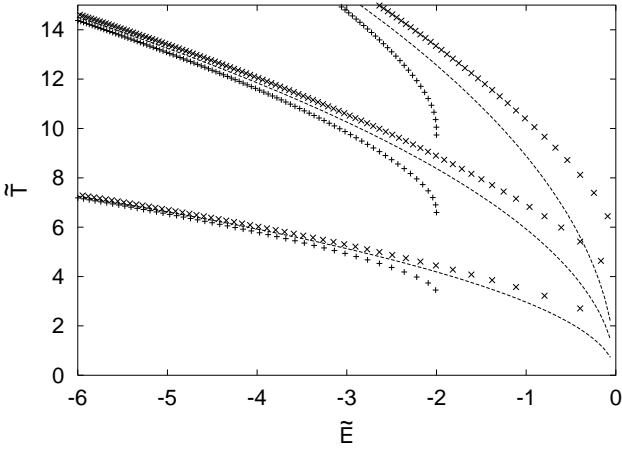}}
  \caption{Bifurcations of non-axial orbits off the downhill (+) and uphill
  ($\times$) orbits in a scaled period vs.~scaled energy plot.
  Bifurcation energies and scaled periods of the bifurcating orbits are
  indicated.  Bifurcations of the downhill orbits only occur
  below the Stark saddle point energy at $\tilde E_{\rm S} = -2$.
  Dashed lines indicate first, second and third multiples of
  $\tilde T=\frac{2\pi}{3}\sqrt{-2\tilde E}$.}
  \label{HeisenbergFig}
\end{figure}

Second, non-axial orbits with winding ratios $z=l/k$ close to one are
created and destroyed at low energies, and their generation and destruction
are closely spaced (see figure~\ref{HeisenbergFig}). Therefore, at low
energies a uniform approximation describing both the creation of a
non-axial orbit from the downhill orbit and its destruction at the uphill
orbit is required. It will be derived in section~\ref{sec:StarkUniform}. As
a prerequisite, to quantitatively assess its necessity, the asymptotic
behaviour of closed-orbit bifurcations at low energies will be investigated
in detail in this section.

Bifurcation energies can be determined from equations (\ref{genEq})
and~(\ref{destEq}). As a winding ratio $z=l/k$ close to one corresponds to
low bifurcation energies and small values of the parameter
$\epsilon=(\tilde E_{\rm S}/\tilde E)^2$, approximate bifurcation energies
can be obtained by expanding the right hand sides of~(\ref{genEq})
and~(\ref{destEq}) in a Taylor series around $\epsilon=0$. We then find the
generation and destruction energies
\begin{equation}
  \label{EBif}
  \begin{split}
  \tilde E_{\rm gen} &= -\frac{\sqrt{3}}{2\sqrt{z-1}}
                -\frac{35}{16\sqrt{3}}\,\sqrt{z-1}
                +{\cal O}\left(z-1\right)^{3/2} \;, \\
  \tilde E_{\rm dest}&= -\frac{\sqrt{3}}{2\sqrt{z-1}}
                +\frac{23}{16\sqrt{3}}\,\sqrt{z-1}
                +{\cal O}\left(z-1\right)^{3/2} \;,
  \end{split}
\end{equation}
for an orbit with a given rational $z$, so that the energy difference
\begin{equation}
  \label{DeltaEBif}
  \Delta \tilde E = \tilde E_{\rm dest}-\tilde E_{\rm gen}
           = \frac{29}{8\sqrt{3}}\,\sqrt{z-1}
            +{\cal O}\left(z-1\right)^{3/2}
\end{equation}
indeed vanishes as $z\to 1$.

\begin{figure}
  \centerline{\includegraphics[width=\columnwidth]{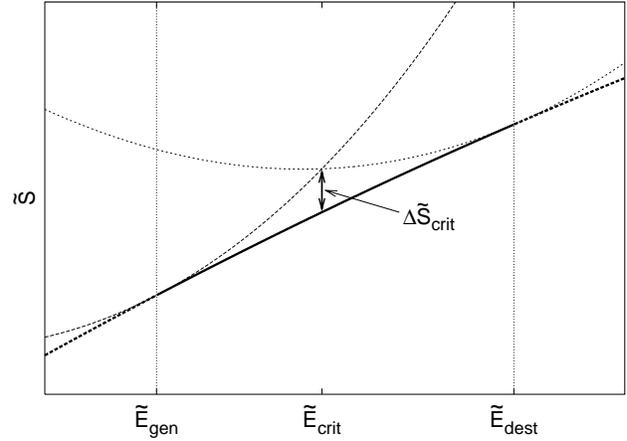}}
  \caption{Schematic illustration of the bifurcation scenario with the
  definition of the critical quantities. Thick solid line: non-axial real
  orbit, thick dashed line: non-axial ghost orbit, thin long-dashed line:
  downhill orbit, thin short-dashed line: uphill orbit. Vertical lines
  indicate the bifurcation energies.}
  \label{CritFig}
\end{figure}

However, the crucial parameter determining whether or not the bifurcations
can be regarded as isolated is not the difference of the bifurcation
energies, but rather the action differences between the non-axial orbit and
the two axial orbits.  For the simple uniform approximation to be
applicable, at any energy at least one of the two action differences must
be large.  In other words (see figure~\ref{CritFig}), at the critical
energy where the actions of the downhill and the uphill orbits are equal,
the difference between their action and the action of the non-axial orbit
must be large.

To begin with, the actions of the downhill and uphill orbits are given by
\begin{equation}
  \label{Sud}
  \begin{split}
  \tilde S_{\begin{array}{c}\mbox{\scriptsize{down}}\\[-1.5ex]
                            \mbox{\scriptsize{up}}\end{array}}
    =& \sqrt{2}\pi(-\tilde E)^{-1/2}
     \pm \frac{3\pi}{4\sqrt{2}}\,(-\tilde E)^{-5/2} \\
     &+   \frac{35\pi}{32\sqrt{2}}\,(-\tilde E)^{-9/2}
     \pm \frac{1155\pi}{512\sqrt{2}}\,(-\tilde E)^{-13/2} \\
     &+   \frac{45045\pi}{8192\sqrt{2}}\,(-\tilde E)^{-17/2}
     +   {\cal O}(-\tilde E)^{-19/2} \;.
  \end{split}
\end{equation}
The critical energy for the orbit $(k,l)$ is characterized by the condition
$k S_{\rm down}=l S_{\rm up}$, so that $S_{\rm down}/S_{\rm up}=z$. We
solve this equation for the critical energy
\begin{equation}
  \label{Ecrit}
  \begin{split}
  \tilde E_{\rm crit} =& -\frac{\sqrt{3}}{2\sqrt{z-1}}
     -\frac{\sqrt{3}}{8}\,\sqrt{z-1}
     -\frac{69\sqrt{3}}{64}\,(z-1)^{3/2} \\
    &+\frac{209\sqrt{3}}{256}\,(z-1)^{5/2}
     +{\cal O}(z-1)^{7/2} \;.
  \end{split}
\end{equation}
As expected, the critical energy lies between the two bifurcation energies.
If it is substituted into (\ref{Sud}), the critical action of the
axial orbits
\begin{equation}
  \label{SaxCrit}
  \begin{split}
  \tilde S_{\rm ax}=&k \tilde S_{\rm down} = l\tilde S_{\rm up}\\ 
    =& \frac{2\pi k}{3^{1/4}}\,(z-1)^{1/4}
     + \frac{3\pi k}{4\cdot 3^{1/4}}\,(z-1)^{5/4} \\
    &- \frac{455\pi k}{576\cdot 3^{1/4}}\,(z-1)^{9/4}
     + {\cal O}(z-1)^{13/4}
  \end{split}
\end{equation}
is obtained.

To determine the critical action of the non-axial orbit, the separation
parameter $c_1$ for this orbit at the critical energy must be calculated
from equation~(\ref{ClosedCond}), which can be rewritten as
\begin{equation}
  \label{closed3}
  \frac{\rho_-\,{\cal K}(m_1)}{\rho_+\,{\cal K}(m_2)} - z = 0 \;.
\end{equation}
Substituting the critical energy (\ref{Ecrit}) into (\ref{closed3}), we find
\begin{equation}
  \label{TauC1}
  \begin{split}
    &\frac{\rho_-\,{\cal K}(m_1)}{\rho_+\,{\cal K}(m_2)} - z 
   = \frac{29(c_1-2)}{24}\,(z-1)^2 \\
    &\qquad+\frac{219\, c_1^2-876\, c_1+526}{144}\,(z-1)^3 \\
    &\qquad+\frac{17951\, c_1^3-107986\, c_1^2+188112\,c_1-76688}
                 {9216}\,(z-1)^4 \\
    &\qquad+{\cal O}(z-1)^5 \;.
  \end{split}
\end{equation}
We now make a power series ansatz for $c_1$, insert it into~(\ref{TauC1})
and solve for the coefficients to find
\begin{equation}
  \label{C1Crit}
  c_1=2+\frac{175}{87}\,(z-1)-\frac{175}{174}\,(z-1)^2
       +{\cal O}(z-1)^3 \;.
\end{equation}
From the knowledge of the critical energy and the separation parameter,
the critical action of the non-axial orbit is found to be
\begin{equation}
  \begin{split}
  \tilde S_{\rm non} =& \frac{2\pi k}{3^{1/4}}\,(z-1)^{1/4}
     + \frac{3\pi k}{4\cdot 3^{1/4}}\,(z-1)^{5/4} \\
    &-\frac{1151\pi k}{576\cdot 3^{1/4}}\,(z-1)^{9/4}
     +{\cal O}(z-1)^{13/4} \;.
  \end{split}
\end{equation}
This result agrees with (\ref{SaxCrit}) in leading and next-to-leading
order. The critical action difference is
\begin{equation}
  \Delta \tilde S_{\rm crit} = \frac{29\pi q}{24\cdot 3^{1/4}}\,(z-1)^{5/4}
       + {\cal O}(z-1)^{9/4}
\end{equation}
with $q=l-k$. If the critical action difference is written as a function of
the critical energy given by~(\ref{Ecrit}), it reads
\begin{equation}
  \label{dSCrit}
  \Delta \tilde S_{\rm crit} = 
   \frac{29\pi q}{32\sqrt{2}}\,(-\tilde E_{\rm crit})^{-5/2}
       + {\cal O}(-\tilde E_{\rm crit})^{-7/2} \;.
\end{equation}
Note that due to delicate cancellations occurring
during the calculation, all terms of the series expansions given above are
needed to obtain the leading-order result (\ref{dSCrit}). Due to these
cancellations, the critical action difference is proportional to a high
power of the energy, so that it vanishes fast at low energies.

\begin{figure}
  \centerline{\includegraphics[width=\columnwidth]{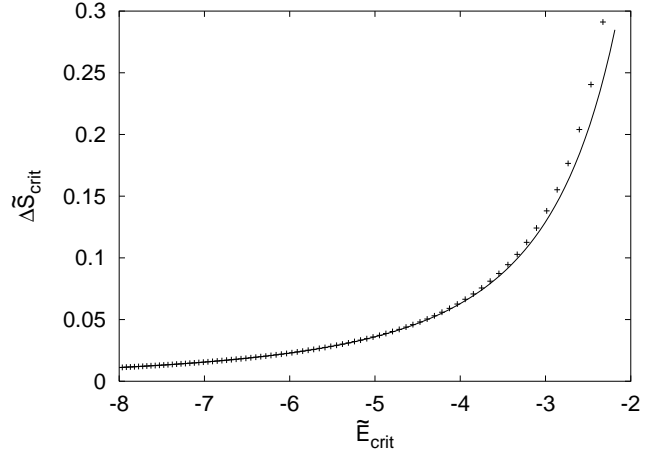}}
  \caption{Numerically calculated critical action differences (crosses) for
  $q=l-k=1$ as
  a function of the critical energies. The solid line gives the
  leading-order result (\ref{dSCrit}).}
  \label{dSCritFig}
\end{figure}

Figure~\ref{dSCritFig} shows numerically computed
critical action differences as a function of the pertinent critical
energies for $q=1$. The leading-order result (\ref{dSCrit}) is indicated by
the solid line. It is accurate even at fairly high energies.

\section{Uniform approximation on the sphere}
\label{sec:StarkUniform}

In the hydrogen atom in an electric field, the bifurcation scenarios are
largely shaped by the rotational symmetry around the electric field axis:
Due to that symmetry, all non-axial orbits occur in one-parameter families,
whereas away from a bifurcation the axial orbits are isolated. A normal
form describing the bifurcation must also possess a rotational
symmetry. Therefore, its configuration space must be at least
two-dimensional. It can be chosen to be defined on a Euclidean plane
$(x,y)$ and reads \cite{Gao97,Shaw98a,Shaw98b,Bartsch03d}
\begin{equation}
  \label{StarkNF}
  \Phi_a(x,y)=\frac14\, r^4 - \frac12 a r^2
\end{equation}
with a real parameter $a$ and $r^2=x^2+y^2$ the distance from the symmetry
centre.  Apart from the trivial stationary point at the origin,
(\ref{StarkNF}) has a ring of stationary points at $r=\sqrt{a}$, which is
real if $a>0$ and imaginary if $a<0$. Thus, (\ref{StarkNF}) describes a
family of non-axial real orbits present for $a>0$ which then contracts onto
an axial orbit and becomes a family of ghost orbits. It therefore
reproduces the elementary bifurcation described in
section~\ref{sec:StarkClosed}.

In terms of the actions $S_{\rm ax}, S_{\rm non}$ and semiclassical
amplitudes ${\cal A}_{\rm ax}$, ${\cal A}_{\rm non}$ of the axial and
non-axial orbits, respectively, we found the uniform
approximation~(\ref{StarkUnif}) corresponding to the normal
form~(\ref{StarkNF}) to be \cite{Bartsch03d}
\begin{equation}
  \label{StarkUnifFinal}
  \Psi(E) = 
    \left[\frac{{\cal A}_{\rm non}}{(1+\ri)}\,I_0
         +\frac{1}{a}\left(a {\cal A}_{\rm ax}
                           +\frac{1+\ri}{\sqrt{2\pi}}\,{\cal A}_{\rm non}
                     \right)
    \right] \re^{\ri S_{\rm ax}} \;,
\end{equation}
where
\begin{equation}
  \label{StarkADef}
  a = \pm 2\sqrt{S_{\rm ax}-S_{\rm non}}
\end{equation}
and
\begin{equation}
  \label{StarkI0}
  I_0=\re^{-\ri a^2/4}
      \left[\frac{1+\ri}{2}-C\left(-\frac{a}{\sqrt{2\pi}}\right)
                         -\ri S\left(-\frac{a}{\sqrt{2\pi}}\right)
      \right]
\end{equation}
with the Fresnel integrals \cite{Abramowitz}
\begin{equation}
  \label{FresnelDef}
  C(x)=\int_0^x \cos\left(\frac\pi 2\,t^2\right) dt \;,\quad
  S(x)=\int_0^x \sin\left(\frac\pi 2\,t^2\right) dt \;.
\end{equation}

We saw in section~\ref{sec:StarkLow} that it is essential at low energies
to construct a uniform approximation capable of simultaneously describing
both bifurcations of a non-axial orbit. As the normal form~(\ref{StarkNF})
describes only one axial orbit and a non-axial orbit bifurcating out of it,
the uniform approximation~(\ref{StarkUnifFinal}) is unsuitable for this
purpose.

A normal form describing both the generation and the destruction of a
non-axial orbit must possess two isolated stationary points corresponding
to the uphill and downhill orbits and a ring of stationary points related
to the family of non-axial orbits. As the normal form parameters are
varied, the ring must branch off the first isolated stationary point and
later contract onto the second. This is impossible if the normal form is to
be defined in a plane, but it is easily achieved if the normal form is
based on a sphere. It can then be chosen in such a way that it has isolated
stationary points at the poles and a ring of stationary points moving from
one pole to the other. A normal form satisfying these requirements is
\begin{equation}
  \label{SNonNF}
  \Phi_{a,b}(\vartheta)=b(\cos\vartheta-a)^2 \;.
\end{equation}
It is given in terms of the polar angle $\vartheta$ on the sphere. Due to
the rotational symmetry, it is independent of the azimuth angle $\varphi$,
and it contains two real parameters $a$ and $b$ needed to match the two
action differences between the three closed orbits.

If, in the vicinity of the poles, $\Phi_{a,b}$ is expanded in terms of the
distance $\rho=\sin\vartheta$ from the polar axis, it reads
\begin{equation}
  \label{SNon0}
  \Phi_{a,b}=b(a-1)^2+ba\left(\frac14\rho^4+\frac{a-1}{a}\rho^2\right)
               + {\cal O}\left(\rho^6\right)
\end{equation}
around $\vartheta=0$, and
\begin{equation}
  \label{SNonPi}
  \Phi_{a,b}=b(a+1)^2-ba\left(\frac14\rho^4+\frac{a+1}{a}\rho^2\right)
               + {\cal O}\left(\rho^6\right)
\end{equation}
around $\vartheta=\pi$, so that in both cases (\ref{SNonNF}) reproduces the
simpler normal form (\ref{StarkNF}) up to re-scaling. From the coefficients
of the quadratic terms in (\ref{SNon0}) and (\ref{SNonPi}) it is apparent
that a ring of stationary points bifurcates off $\vartheta=0$ at $a=1$, and
off $\vartheta=\pi$ at $a=-1$.

These conclusions are confirmed by a discussion of the full normal form
(\ref{SNonNF}). The ring of stationary points is located at the polar angle
$\vartheta_{\rm non}$ satisfying 
\begin{equation} 
  \label{SNonTheta}
  \cos\vartheta_{\rm non}=a\;.
\end{equation}
This angle qualitatively corresponds to the starting and returning angle of
the non-axial orbit in that it describes its motion from the downhill orbit
at $\vartheta=0$ to the uphill orbit at $\vartheta=\pi$. Quantitatively,
$\vartheta_{\rm non}$ cannot directly be identified with the classical
angle $\vartheta$ because the energy-dependence of the normal form
parameters $a$ and $b$ is fixed by matching the action differences rather
than the angles.  If $-1\le a\le 1$, the angle $\vartheta_{\rm non}$
defined by (\ref{SNonTheta}) is real. If $a>1$, $\vartheta_{\rm non}$ is
purely imaginary, corresponding to a ghost orbit at energies below~$\tilde
E_{\rm gen}$. For $a<-1$, $\vartheta_{\rm non}$ is of the form
$\vartheta_{\rm non}=\pi+\ri\alpha$, as is characteristic of a ghost orbit
at energies higher than $\tilde E_{\rm dest}$.

\begin{figure}
  \centerline{\includegraphics[width=\columnwidth]{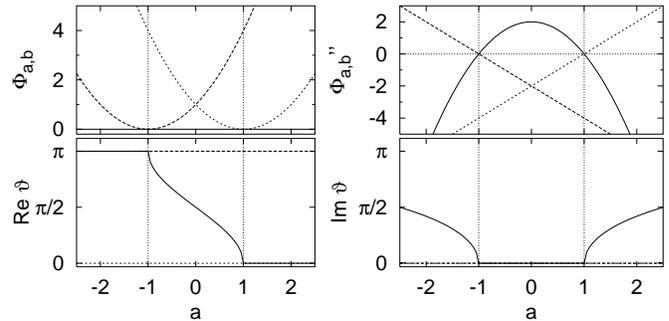}}
  \caption{Bifurcation scenario described by the normal form
  (\ref{SNonNF}) for $b=1$. Location $\vartheta$ of the stationary points,
  stationary values of the normal form $\Phi_{a,b}$ and stationary values
  of its second derivative. Solid line: stationary point at $\vartheta_{\rm
  non}$, long-dashed line: stationary point at $\vartheta=\pi$,
  short-dashed line: stationary point at $\vartheta=0$.}
  \label{StarkNFFig}
\end{figure}
\begin{figure}
  \centerline{\includegraphics[width=\columnwidth]{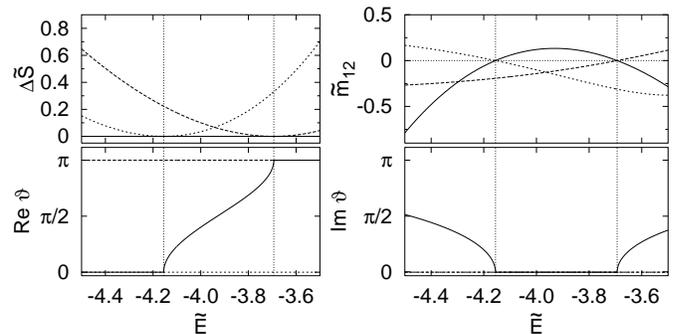}}
  \caption{Orbital parameters for the bifurcations of the (20,21) non-axial
  orbit.  Solid line: non-axial orbit $(20,21)$, long-dashed line: 21st
  repetition of the
  uphill orbit, short-dashed line: 20th repetition of the downhill
  orbit. The vertical lines
  indicate the bifurcation ener\-gies $\tilde E_{\rm gen}=-4.155542$ 
  and $\tilde E_{\rm dest}=-3.693431$, respectively.}
  \label{StarkBif20Fig}
\end{figure}

The bifurcation scenario described by the normal form (\ref{SNonNF}) is
summarized in figure \ref{StarkNFFig}. For $b=1$, the
stationary values of (\ref{SNonNF}), the complex polar angles $\vartheta$
where stationary points occur, and the second derivative of $\Phi_{a,b}$ with
respect to $\vartheta$ are shown as functions of $a$. Due to the
rotational symmetry, the latter is well defined even at the poles.

Figure \ref{StarkNFFig} should be compared to figure \ref{StarkBif20Fig},
which displays the orbital data pertinent to an actual bifurcation
scenario. The qualitative agreement between figure \ref{StarkNFFig} and
figure \ref{StarkBif20Fig} is obvious, so that (\ref{SNonNF}) can be seen
to describe the bifurcations correctly. (Note that $a$ is a decreasing
function of $E$.)

As the normal form (\ref{SNonNF}) has been chosen such as to vanish at the
non-axial stationary point $\vartheta_{\rm non}$, the pertinent uniform
approximation reads
\begin{equation}
  \label{SNonUnif}
  \Psi(E) = I(a,b) \,\re^{\ri S_{\rm non}(E)} \;,
\end{equation}
where $S_{\rm non}(E)$ denotes the action of the non-axial orbit and 
\begin{equation}
  \label{SNonInt}
  I(a,b)=\int_0^\pi d\vartheta \int_0^{2\pi} d\varphi \,
           p(\vartheta)\, \re^{\ri\Phi_{a,b}(\vartheta)}
\end{equation}
with an amplitude function $p(\vartheta)$ to be specified below.

The connection with closed-orbit theory is established with the help
of a stationary-phase approximation to the integral~(\ref{SNonInt}). If
$|a|>1$, the only real stationary points of $\Phi_{a,b}$ are located at the
poles, and the asymptotic approximation to (\ref{SNonUnif}) reads
\begin{equation}
  \label{SNonInt1}
  \begin{split}
  \Psi(E)\approx 
      &-\frac{\pi\ri}{b(1-a)}\,p(0)\re^{\ri b(1-a)^2+\ri S_{\rm non}}  \\
      &-\frac{\pi\ri}{b(1+a)}\,p(\pi)\re^{\ri b(1+a)^2+\ri S_{\rm non}} \;.
  \end{split}
\end{equation}
If $|a|<1$, the contribution
\begin{equation}
  \label{SNonInta}
  \Psi(E)|_{\vartheta_{\rm non}} = 2\pi\sqrt{\frac{\pi\ri}{b}}
	\,p(\vartheta_{\rm non})\,\re^{\ri S_{\rm non}}
\end{equation}
has to be added to (\ref{SNonInt1}).
These asymptotic expressions should reproduce the simple closed-orbit
theory formulae
\begin{equation}
  {\cal A}_{\rm down}\re^{\ri S_{\rm down}} +
  {\cal A}_{\rm up}\re^{\ri S_{\rm up}}
\end{equation}
and
\begin{equation}
  {\cal A}_{\rm non}\re^{\ri S_{\rm non}} \;,
\end{equation}
respectively. We thus obtain the conditions
\begin{equation}
  \label{SNonS}
  \begin{split}
    \Delta S_{\rm down}&=S_{\rm down}-S_{\rm non}=b(1-a)^2\;,\\
    \Delta S_{\rm up}&=S_{\rm up}-S_{\rm non}=b(1+a)^2 \;,
  \end{split}
\end{equation}
and
\begin{equation}
  \label{SNonA}
  \begin{gathered}
    {\cal A}_{\rm down}=-\frac{\pi\ri}{b(1-a)}\,p(0) \;,\qquad
    {\cal A}_{\rm up}=-\frac{\pi\ri}{b(1+a)}\,p(\pi) \;,\\
    {\cal A}_{\rm non}=2\pi\sqrt{\frac{\pi\ri}{b}}\,p(\vartheta_{\rm non})
  \end{gathered}
\end{equation}
connecting the actions and amplitudes of the downhill, uphill and non-axial
orbits to the normal form parameters $a$ and $b$ and to the values of the
amplitude function $p(\vartheta)$.

Equation (\ref{SNonS}) can be solved for the normal form parameters to
yield
\begin{equation}
  \label{SNona}
  a=\frac{\sqrt{\Delta S_{\rm up}}\mp\sqrt{\Delta S_{\rm down}}}
         {\sqrt{\Delta S_{\rm up}}\pm\sqrt{\Delta S_{\rm down}}}
\end{equation}
and
\begin{equation}
  \label{SNonb}
  b=\frac{\Delta S_{\rm down}}{(1-a)^2} = \frac{\Delta S_{\rm up}}{(1+a)^2}\;,
\end{equation}
where the upper or lower sign has to be chosen if the non-axial orbit is
real or complex, respectively.

If we set $p(\vartheta)$ equal to a constant in (\ref{SNonA}),
we obtain the constraints
\begin{equation}
  \label{SNonConst}
  \begin{split}
  \frac{1+\ri}{2}\,\sqrt{\frac{b}{2\pi}}\,{\cal A}_{\rm non}
    &= -b(1-a)\,{\cal A}_{\rm down} \\
  \frac{1+\ri}{2}\,\sqrt{\frac{b}{2\pi}}\,{\cal A}_{\rm non}
    &= -b(1+a)\,{\cal A}_{\rm up}
  \end{split}
\end{equation}
the amplitudes must satisfy close to the individual
bifurcations. Numerically, (\ref{SNonConst}) can be verified to hold in the
neighbourhood of the bifurcations.

To obtain a complete uniform approximation, we make the ansatz
\begin{equation}
  \label{SNonpDef}
  p(\vartheta)=p_0+p_1(\cos\vartheta-a)+p_2(\cos\vartheta-a)^2
\end{equation}
for the amplitude function.
The coefficients are given by
\begin{equation}
  \label{SNonCoef}
  \begin{gathered}
    p_0=p(\vartheta_{\rm non}) \;, \\
    p_1=\frac{(1+a)^2p(0)-4a\,p(\vartheta_{\rm non})-(1-a)^2p(\pi)}
              {2(1-a^2)} \;, \\
    p_2=\frac{(1+a)p(0)-2p(\vartheta_{\rm non})+(1-a)p(\pi)}
              {2(1-a^2)}
  \end{gathered}
\end{equation}
with the values $p(0)$, $p(\pi)$ and $p(\vartheta_{\rm non})$ determined
from (\ref{SNonA}). Due to (\ref{SNonConst}), all coefficients are
finite everywhere.

With this choice, the uniform approximation (\ref{SNonUnif}) reads
\begin{equation}
  \label{SNonUFinal}
  \Psi(E)=p_0 I_0 + p_1 I_1 + p_2 I_2
\end{equation}
with the integrals
\begin{equation}
  \begin{split}
    I_0&=\int_0^{2\pi} d\varphi \int_0^\pi d\vartheta \sin\vartheta  \,
          \re^{\ri b(\cos\vartheta-a)^2} \\
       &=2\pi\sqrt{\frac{\pi}{2b}}
         \left(C\left(\sqrt{\frac{2b}{\pi}}(1-a)\right)+
               \ri S\left(\sqrt{\frac{2b}{\pi}}(1-a)\right)\right.\\
       &\phantom{=2\pi\sqrt{\frac{\pi}{2b}}\,}\left.
               +\,C\left(\sqrt{\frac{2b}{\pi}}(1+a)\right)+
               \ri S\left(\sqrt{\frac{2b}{\pi}}(1+a)\right)
         \right)
  \end{split}
\end{equation}
in terms of the Fresnel integrals (\ref{FresnelDef}),
\begin{equation}
  \begin{split}
    I_1&=\int_0^{2\pi} d\varphi \int_0^\pi d\vartheta \sin\vartheta  \,
          (\cos\vartheta-a)\,\re^{\ri b(\cos\vartheta-a)^2} \\
       &=\frac{\pi\ri}{b}\left(\re^{\ri b(1+a)^2}-\re^{\ri b(1-a)^2}\right)
  \end{split}
\end{equation}
and
\begin{equation}
  \begin{split}
    I_2&=\int_0^{2\pi} d\varphi \int_0^\pi d\vartheta \sin\vartheta\,
          (\cos\vartheta-a)^2\,\re^{\ri b(\cos\vartheta-a)^2} \\
       &= -\ri \frac{d}{d b} I_0 \\
       &= -\frac{\pi\ri}{b}
           \left((1-a)\,\re^{\ri b(1-a)^2}+(1+a)\,\re^{\ri b(1+a)^2}\right)
          +\frac{\ri}{2b}I_0 \;.
  \end{split}
\end{equation}
All integrals are finite as they stand, they do not require any
regularization. This somewhat atypical behaviour can be traced back to the
fact that the domain of integration in this case is compact.

\begin{figure}
  \centerline{\includegraphics[width=\columnwidth]{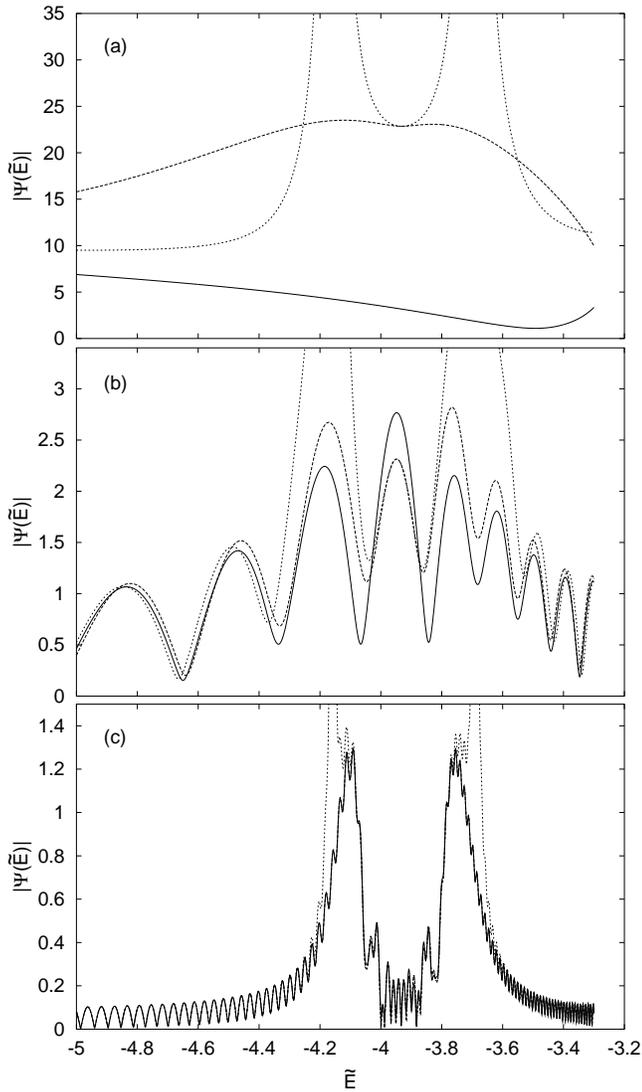}}
  \caption{Uniform approximation for the generation and destruction of the
  (20,21) non-axial orbit for the electric field strengths (a) $F=10^{-2}$,
  (b) $F=10^{-6}$, (c) $F=10^{-10}$. Solid lines: Uniform approximation
  (\ref{SNonUFinal}), long-dashed lines: uniform approximation
  (\ref{StarkUnifFinal}) for isolated bifurcations, short-dashed lines:
  isolated-orbits approximation.}
  \label{Unif2Fig}
\end{figure}

As an example, figure~\ref{Unif2Fig} compares the uniform approximation
(\ref{SNonUFinal}) for the orbit (20,21) and three different electric field
strengths to the isolated-orbits approximation and to the uniform
approximation (\ref{StarkUnifFinal}) for isolated bifurcations. For the
latter, the non-axial orbit and the axial orbit for which the action
difference is smaller are combined into the uniform approximation, where\-as
the second axial orbit is treated as isolated.

For the high electric field strength $F=10^{-2}$ in
figure~\ref{Unif2Fig}(a), in the energy range shown none of the uniform
approximations reaches the asymptotic region where they agree with the
isolated-orbits approximation on either side of the bifurcations. Between
the two bifurcations, the isolated-bifurcations approximation appears just
to arrive at its asymptotic behaviour before the second bifurcation is
felt. The behaviour of the uniform approximation on the sphere is completely
different there.

If the field strength is decreased, both uniform approximations reach their
asymptotic regions faster. For $F=10^{-6}$, they approach the
isolated-orbits approximation at roughly equal pace on both sides of the
bifurcations. Between the bifurcations, the isolated-bifurcations
approximation now clearly attains its asymptotic behaviour, where\-as the
uniform approximation on the sphere exhibits qualitatively similar behaviour,
but differs significantly from the other two in quantitative terms. It
requires the even lower field strength of $F=10^{-10}$ used in
figure~\ref{Unif2Fig}(c) until both uniform approximations agree with the
isolated-orbits result in the region between the bifurcations.

The uniform approximation on the sphere was constructed because in the
energy region between the bifurcations the isolated-bifurcations
approximation must fail for low energies and high field strengths. However,
it turns out to reach its asymptotic agreement with the isolated-orbits
results considerably more slowly than the isolated-bifurcations
approximation. In cases where the latter reproduces the isolated-orbits
results between the bifurcations, of course, no further uniformization is
required.

\section{Semiclassical Stark spectra}
\label{sec:StarkScl}

More information on the virtue of the uniform approximation on the sphere
can be obtained from a calculation of actual semiclassical spectra. To this
end we calculated semiclassical photo-absorption spectra for the hydrogen
atom in an electric field based on closed-orbit theory and including
uniform approximations. To low-resolution, a semiclassical spectrum can be
obtained if the contributions of orbits up to a maximum period $\widetilde
T_{\rm max}$ are included in the closed-orbit sum~(\ref{resOscGen}).

We will present results for the photoexcitation of the $1s$ ground state of
the hydrogen atom in an electric field of strength $F=10^{-8}\,{\rm a.u.}
\hat = 51.4\,{\rm V/cm}$. The exciting photon is assumed to be linearly
polarized along the field axis.
In the closed-orbit summation, the uniform approximation~(\ref{SNonUFinal})
based on the sphere will be used if the action differences between a
non-axial orbit and both the downhill and uphill orbits it bifurcates out
of are smaller than $2\pi$. In the other cases, we revert to the simple
uniform approximation~(\ref{StarkUnifFinal}) or the isolated-orbits
approximation, as appropriate.

\begin{figure}
  \centerline{\includegraphics[width=\columnwidth]{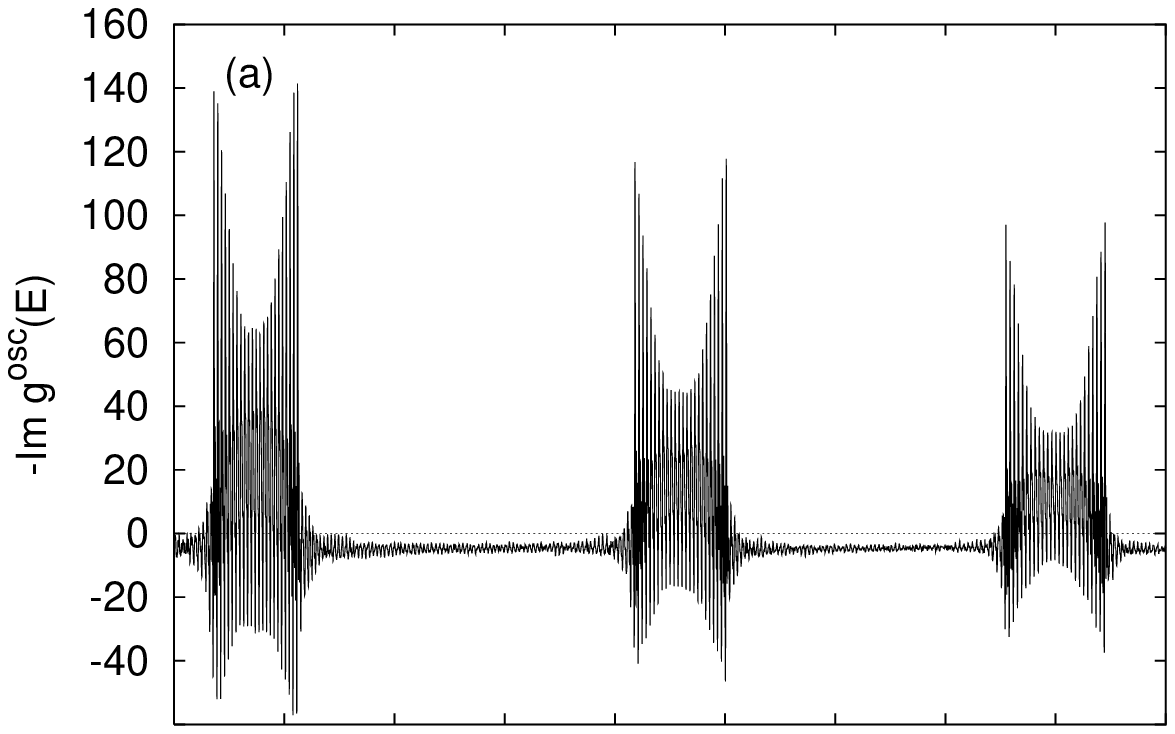}}
  \centerline{\includegraphics[width=\columnwidth]{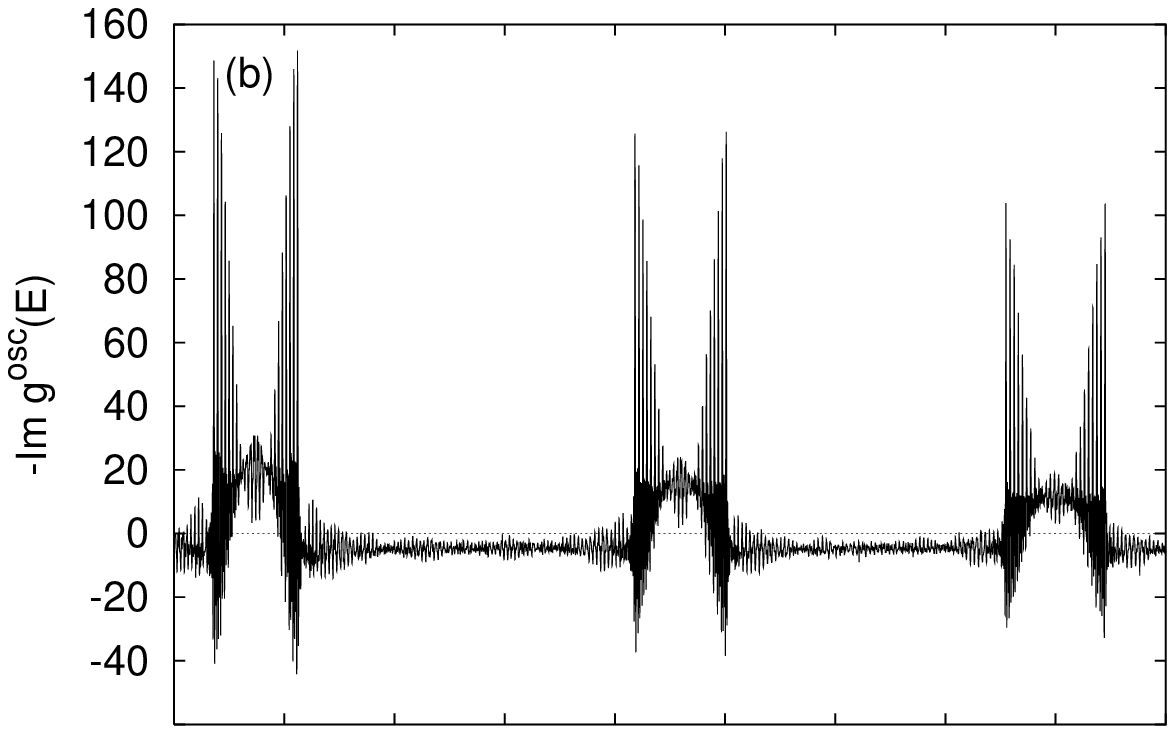}}
  \centerline{\includegraphics[width=\columnwidth]{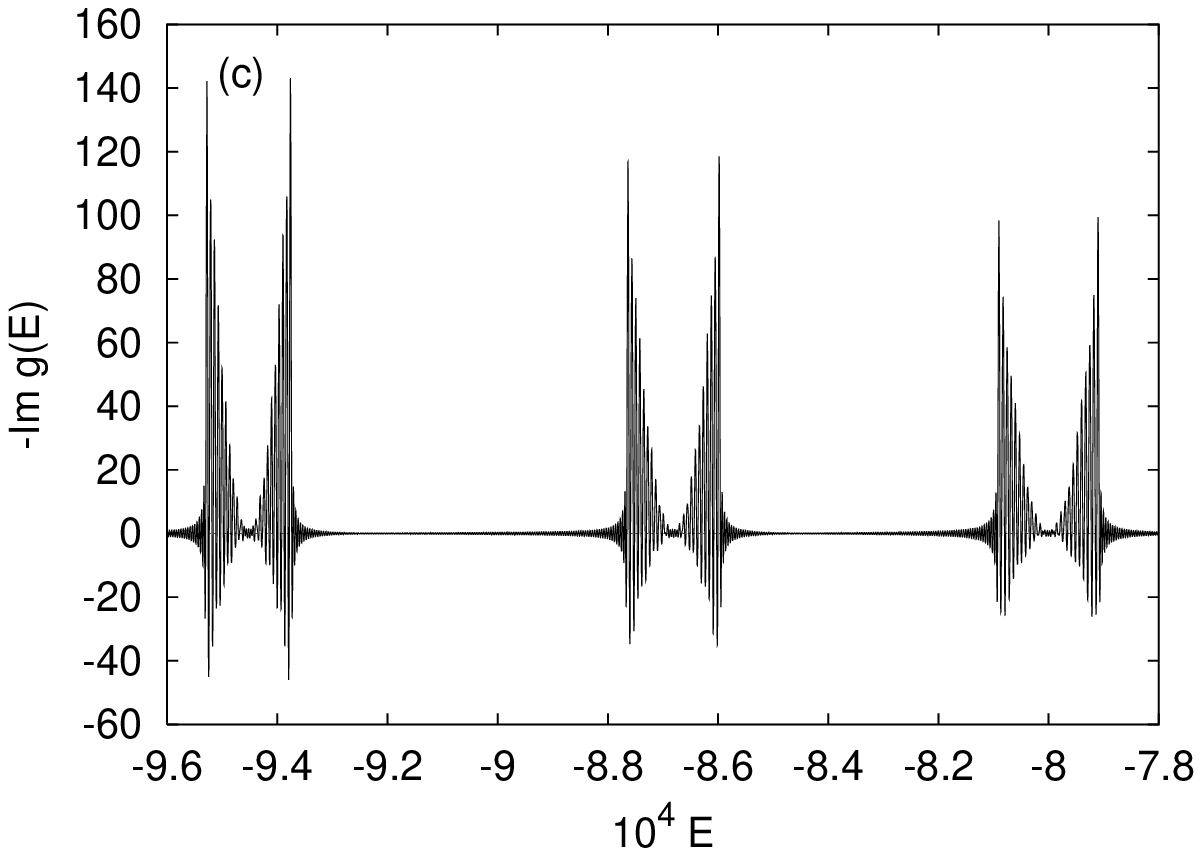}}
  \caption{Low-resolution photo-absorption spectrum: (a) semiclassical
  spectrum not using and (b) using the uniform approximation on the sphere,
  (c) smoothed quantum spectrum.  The scaled cut-off time is $\widetilde
  T_{\rm max}=30$.}
  \label{StarkT30LoFig}
\end{figure}

Figure~\ref{StarkT30LoFig} shows a comparison between the semiclassical
spectra obtained from the simple uniform approximation only
(figure~\ref{StarkT30LoFig}(a)), the
closed-orbit summation with the uniform approximation based on the sphere
(figure~\ref{StarkT30LoFig}(b)), and an appropriately smoothed quantum
spectrum (figure~\ref{StarkT30LoFig}(c)).  Note that only the oscillatory
part of the semiclassical spectra is plotted, so that the base line is
slightly shifted away from zero.  This shift is absent in the quantum
spectrum.

It has already been found in \cite{Bartsch03d} that, if only the simple
uniform approximation~(\ref{StarkUnif}) is used, the spectral lines close
to the centres of the $n$-manifolds tend to have considerable strengths
although they are negligibly weak in the quantum spectrum. If the uniform
approximation on the sphere is used, on the one hand the line strengths
develop a second maximum at the centre of a manifold. On the other hand,
however, the absolute intensity of the central lines is considerably
reduced, so that the overall intensity distribution within an $n$-manifold
is reproduced much better than with the isolated-bifurcations approximation
only.

\begin{figure}
  \centerline{\includegraphics[width=\columnwidth]{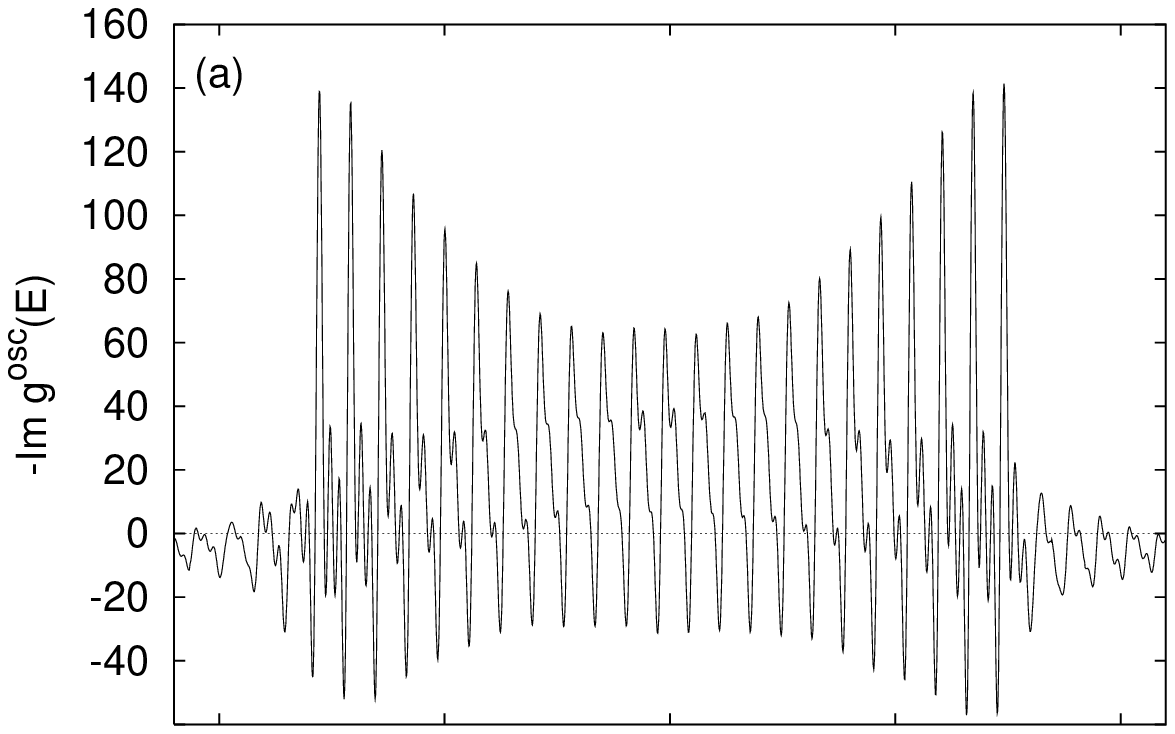}}
  \centerline{\includegraphics[width=\columnwidth]{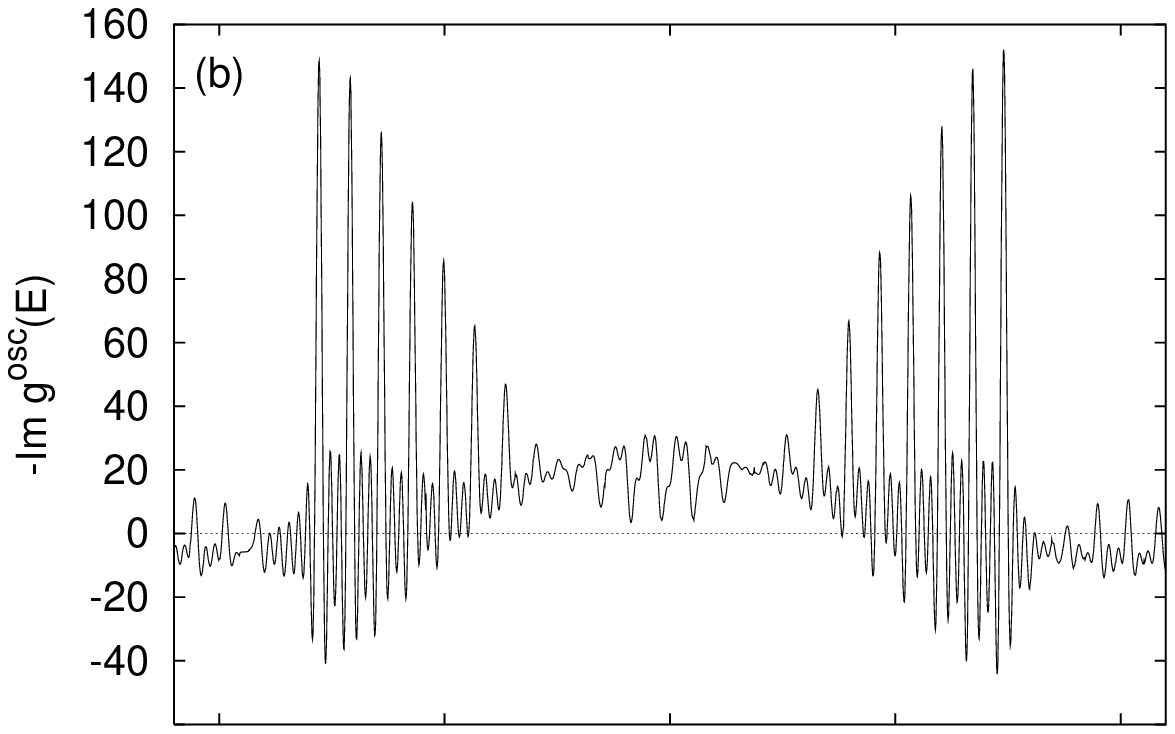}}
  \centerline{\includegraphics[width=\columnwidth]{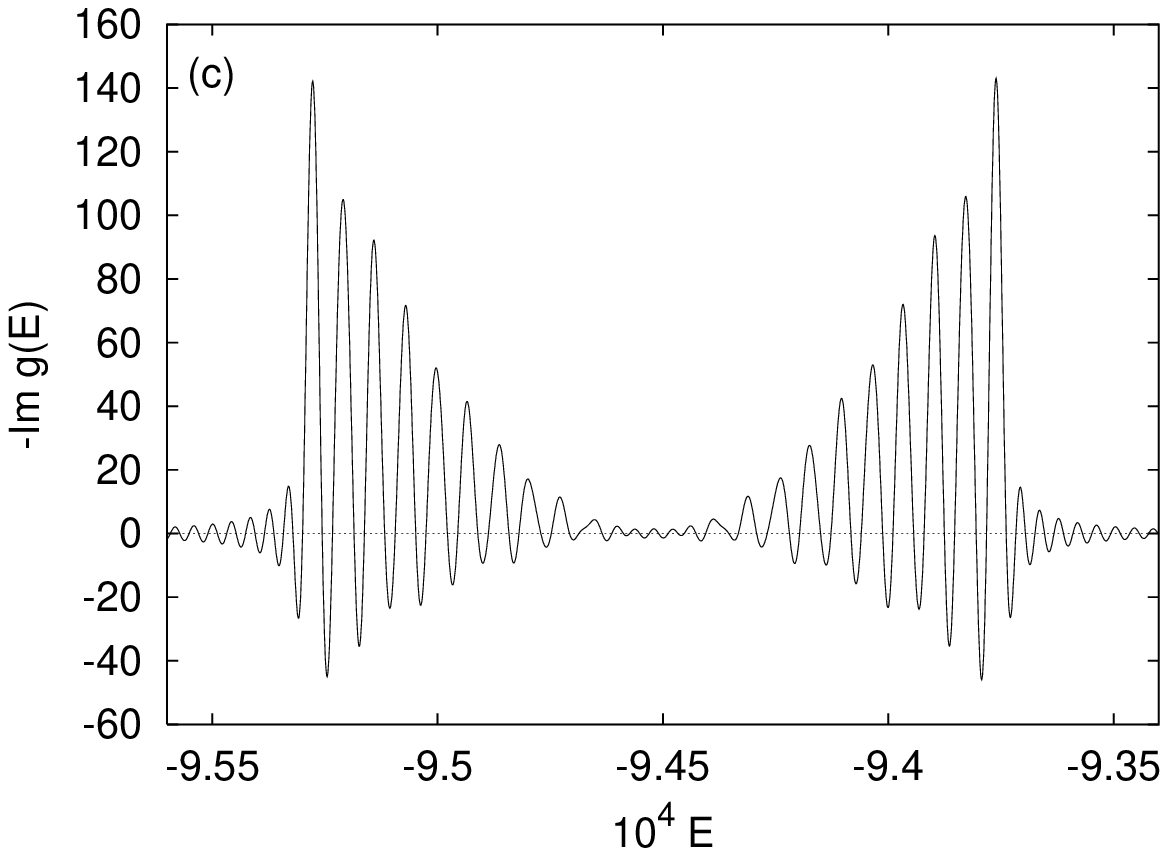}}
  \caption{The manifold $n=23$ from the spectra shown in
  figure~\ref{StarkT30LoFig}. (a) Semiclassical spectrum not using and (b)
  using the uniform approximation on the sphere, (c) smoothed quantum
  spectrum.}
  \label{StarkLoMFig}
\end{figure}

These observations can be confirmed if one of the $n$-manifolds is viewed
in more detail.  Figure~\ref{StarkLoMFig} compares the low-resolution
semiclassical spectra of the manifold $n=23$, which is the lowest
$n$-manifold included in the calculation, to the quantum spectrum. In the
spectrum shown in figure~\ref{StarkLoMFig}(a), which uses the simple
uniform approximation~(\ref{StarkUnifFinal}) only, the lines at the edges
of the manifold are the strongest, and the line strengths decrease gently
toward the centre. The line shapes are slightly asymmetric, although the
asymmetry is weaker than at higher energies \cite{Bartsch03d}.
In the spectrum obtained using the uniform approximation on the sphere 
(figure~\ref{StarkLoMFig}(b)), the strong outer lines of the manifold 
are virtually identical to those in the previous spectrum, but the central 
lines are distorted.
On the other hand, again, as the central lines are weaker in the spectrum 
of figure~\ref{StarkLoMFig}(b) than in figure~\ref{StarkLoMFig}(a), the 
distribution of intensities within the manifold is described more 
accurately if the uniform approximation on the sphere is used.

In addition to the calculation of low-resolution semiclassical spectra, we
can use the techniques of \cite{Bartsch02a,Bartsch03d} to compute
individual spectral lines.  The results do not depend strongly on the
choice of the uniform approximations; the high-resolution spectra obtained
from the spectrum in figure~\ref{StarkT30LoFig}(b) which use the uniform
approximation on the sphere (see \cite{Bartsch02}) are virtually identical
to those presented in \cite{Bartsch03d} that apply the simple uniform
approximation (figure~\ref{StarkT30LoFig}(a)) only.

The uniform approximation on the sphere, although it satisfies the correct
asymptotic conditions and is well adapted to the bifurcation scenario
qualitatively, in quantitative terms does not yield absolutely precise
results.  The problem is probably connected to the non-trivial spherical
topology of the configuration space of the uniform approximation: The
construction of uniform approximations of the type~(\ref{StarkUnif}) relies
on the assumption that the unknown phase function in the diffraction
integral can be mapped to a certain standard form by a suitable coordinate
transformation. The fundamental theorems of catastrophe theory
\cite{Castrigiano93,Poston78} guarantee that this is possible locally.  In
a Cartesian configuration space and close to a bifurcation, all stationary
points of the normal form are close to the origin, so that coordinate
regions away from the origin can, in the spirit of the stationary-phase
approximation, be assumed not to contribute to the diffraction integral.  A
local mapping of the function in the exponent to the normal form and a
local approximation to the amplitude function by means of a Taylor series
thus suffice to obtain good results. By contrast, on the spherical
configuration space a global approximation to the unknown functions must be
sought because the stationary points are distributed across the sphere:
There is a stationary point at each pole. In the case at hand, we tried
replacing the $(\cos\vartheta-a)^2$-term in the amplitude function
(\ref{SNonpDef}) with $\sin\vartheta$, which also gives a rough
approximation to the actual amplitude function.  The alteration has no
significant effect on the result.  The construction of a quantitatively
more reliable uniform approximation will presumably require a deeper global
understanding of the bifurcation scenarios.  It remains a worthwhile
challenge for future research to devise an easy-to-apply technique for the
construction of uniform approximations that takes full account of the
classical phase space topology.

\section{Conclusion}
\label{sec:Conc}

We have discussed the bifurcations of closed orbits in the hydrogen atom in
an electric field. At low scaled energies a uniform semiclassical
approximation is required that simultaneously describes both the generation
and the destruction of a non-axial orbit. Its construction was carried out.
Beyond its significance to the Stark system, it is of interest for reasons
of principle because it uses a sphere instead of a Euclidean plane as its
configuration space and thus is the first uniform approximation introduced
in the literature that relies on a topologically non-trivial configuration
space. It could be demonstrated that uniform approximations with a
non-Euclidean configuration space can indeed be constructed and allow for
the semiclassical calculation of significantly improved photo-absorption
spectra of the hydrogen atom in an electric field.


\end{document}